# Dislocation Slip or Phase Transformation Lead to Room-Temperature Plasticity in Diamond: Comment on "Plastic Deformation of Single-Crystal Diamond Nanopillars"

*Yeqiang Bu, Peng Wang, Anmin Nie\*, and Hongtao Wang\**

Dr. Yeqiang Bu, Prof. Hongtao Wang
Center for X-Mechanics, Institute of Applied Mechanics, Zhejiang University, Hangzhou 310027, China
E-mail: htw@zju.edu.cn

Dr. Peng Wang
Materials Genome Institute, Shanghai University, Shanghai 200444, China

Prof. Anmin Nie
Center for High Pressure Science, State Key Laboratory of Metastable Materials Science and Technology, Yanshan University, Qinhuangdao 066004, China
E-mail: anmin@ysu.edu.cn

**Abstract:** Despite decades of extensive research on mechanical properties of diamond, much remains to be understood in term of plastic deformation mechanisms due to the poor deformability at room temperature. In a recent work in *Advanced Materials*, it was claimed that room-temperature plasticity occurred in <001>-oriented single-crystal diamond nanopillars based on observation of unrecovered deformation inside scanning electron microscope. The plastic deformation was suggested to be mediated by a phase transition from $sp^3$ carbon to an O8-carbon phase by molecular dynamics simulations. By comparison, our *in-situ* transmission electron microscopy study reveals that the room-temperature plasticity can be carried out by dislocation slip in both <100> and <111>-oriented diamond nanopillars. The brittle-to-ductile transition is highly dependent on the stress state. We note that the surface structure may play a significant role in the deformation mechanisms as the incipient plasticity always occurs from the surface region in nanoscale diamonds.

**Main Text:**

Diamond is considered to be the hardest crystalline material with exceptional high strength for the strong C-C covalent bonds. [1, 2] It is also known by the extreme brittleness originated from the bond stiffness. [3] Whether room-temperature plasticity exists in diamond has been debated for decades due to the lack of direct evidence, especially *in-situ* observations.

In a recent work published in *Advanced Materials* in 2020, [4] room-temperature plastic deformation was claimed to be observed in <100>-oriented diamond nanopillars with diameters smaller than ~ 25 nm. The nanopillars were fabricated from the bulk diamond by an inductively-coupled-plasma reactive ion etching (RIE) process using oxygen and argon as chemically active and energetic ion bombardment species. The bending experiments were performed *in situ* inside a scanning electron microscope (SEM) by focusing the electron beam onto the diamond nanopillars. The as-resulted electrostatic force from the injected charges led to deflection of the nanopillars. Permanent deformation was observed in the <100>-oriented nanopillars with smaller dimeters, while the <111>-oriented ones exhibited brittle fracture. Further molecular dynamic (MD) simulations were conducted to show that this plastic deformation in diamond nanopillar might be mediated by a phase transition from $sp^3$ carbon to an O8-carbon phase. As understanding on the room-temperature plasticity in diamond is very limited, we briefly comment on Regan *et al.*'s work [4] based on our findings that may draw more attention to this field; however, our intention is not criticism.

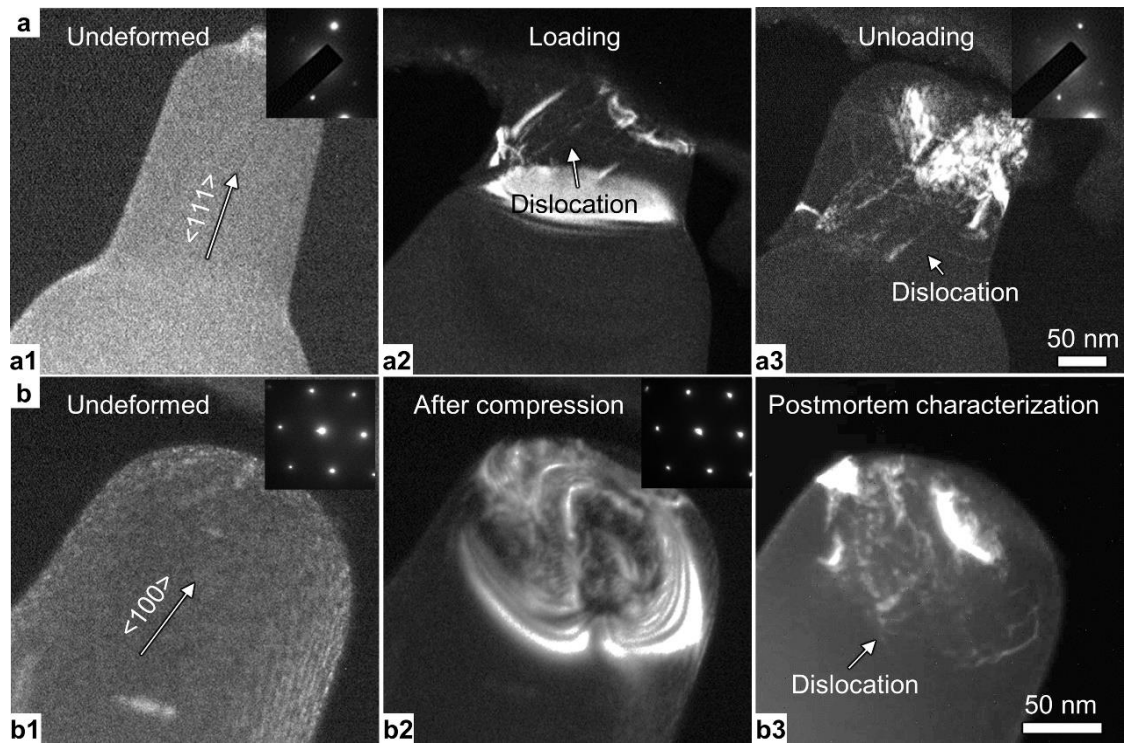

**Figure 1.** Room temperature plastic deformation of single-crystal diamond nanopillars under *in-situ* compression. (a) Dark-field TEM images of a <111>-oriented ~200 nm-sized diamond nanopillar (a1) before brought into contact with, (a2) when compressed against and (a3) after released from the indenter. (b) Dark-field TEM images of a <100>-oriented ~300 nm-sized diamond nanopillar (b1) before brought into contact with and (b2) after released from the indenter. (b3) The postmortem characterization of the same diamond nanopillar in (b2) reveals the dislocation lines. Insets are corresponding selective area electron diffraction (SAED) patterns.

Recently, we performed *in-situ* transmission electron microscopy (TEM) compression on sub-micron-sized diamond pillars at room temperature. Extensive dislocation plasticity has been observed in both <111> and <100>-oriented pillars under compression, as shown in Figs. 1a and b, respectively. Dark-field TEM imaging method was performed to visualize dislocation lines with high contrast. In this operation mode, dislocation lines had bright contrast. The uniform contrast in Fig. 1a1 showed no pre-exist defects inside a <111>-oriented ~200 nm-sized diamond nanopillar. Dislocation lines were observed to be nucleated from the contact region during the *in-situ* compression, as indicated by the arrow in Fig. 1a2. After released, the diamond nanopillar changed its geometry permanently (Fig. 1a3). The experimental results are similar for <100>-oriented diamond nanopillars, as shown in Fig. 1b. No phase transformation can be identified by either TEM imaging or diffraction. It is noted that the large hydrostatic stress component of compression is the key to suppressing catastrophic brittle fracture. For uniaxial compression, the stress ($\sigma$) is decomposed into a hydrostatic pressure ($p$) and a pure shear ($\tau$) with $p = -1/3\sigma$ and $\tau = 1/2\ \sigma$. The resolved normal stress on any lattice plane is no greater than zero. Therefore, no driven force is provided for crack nucleation and propagation, *i.e.* the cleavage process, under compression. All deformation under compression is accommodated by dislocation slips driven by the shear stress component.

On the other hand, our *in-situ* bending experiments on high-aspect-ratio diamond nanopillars reveal that large elastic deformation before cleavage is the only observed mechanical behavior, regardless of orientation and diameters. [5] The <100>-oriented nanopillars with diameter less than 60 nm exhibit highest elastic tensile strain (13.4%) and tensile strength (125 GPa). These values are comparable with the theoretical elasticity and Griffith strength limits of diamond, respectively. For the <111>-oriented nanopillars with diameter less than 30 nm, the maximum achieved tensile strain is lowered to 9.4%, close to a recent work by Banerjee *et al.* published in *Science* in 2018.[2] Our results clearly indicate both the strong elastic anisotropy and the lack of plasticity when diamond nanopillars subjected to bending.

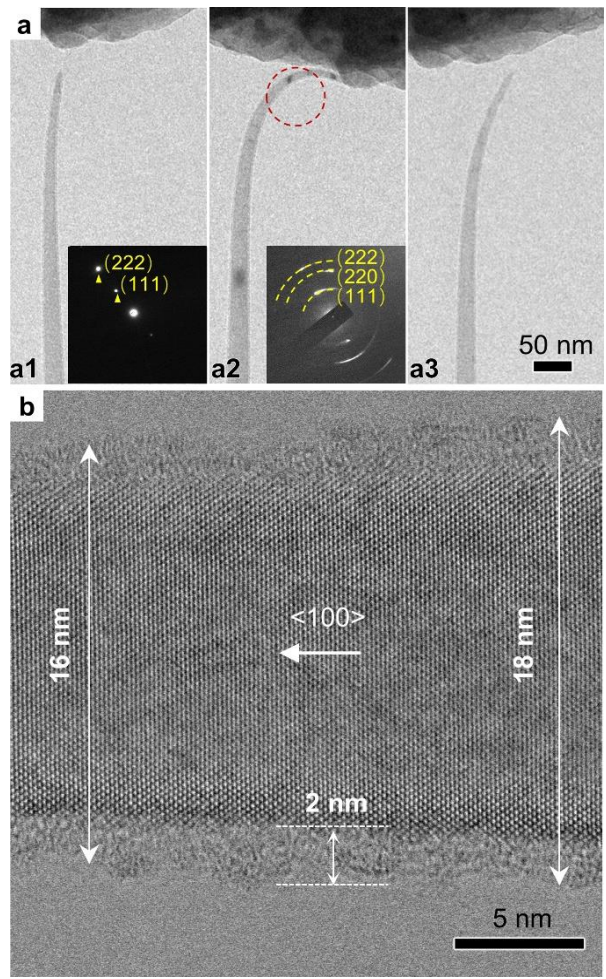

**Figure 2.** (a) The mechanical response of the diamond nanopillar during an *in-situ* bending experiment. Bright field TEM images of a <100>-oriented diamond nanopillar with high aspect ratio (a1) before brought into contact with, (a2) when pushed against and (a3) after released from the indenter. Insets to (a1) and (a2) are the corresponding SAED patterns. (b) High-resolution scanning TEM image of ~20 nm-sized diamond nanopillar.

It is noted that the free surface may impose strong effect on the deformation of diamond nanopillars with diameters less than 20 nm. Figure 2a shows the *in-situ* bending of a <100>-oriented diamond nanopillar. About 90° deflection was observed in the front region with diameter around 20 nm (Fig. 2a2). Inset to Fig. 2a2 shows a ring-like SAED pattern due to the continuous rotation of lattice planes in the bending region. No extra rings or spots from a new phase showed up. After unloading, permanent deflection was identified (Fig. 2a3). We note that the residual deformation is not caused by the plastic deformation in diamond, but the amorphous surface layer as shown in Fig. 2b. The nanopillar was fabricated by focused ion beam (FIB) milling from natural type Ia diamond with a current of 0.5 nA under a voltage of 30 kV. The diameter was gradually reduced from 200 nm to less than 20 nm in the front end by Argon plasma thinning for a total of eight hours by decreasing voltages from 2.0 kV to 1.5, 1.0, and finally 0.8 kV. This step also removed possible contamination and amorphous carbon unintentionally introduced during the FIB milling. After such a dedicated cleaning process, there still remains a ~2 nm-thick amorphous carbon surface layer (Fig. 2b), occupying approximately 36% in volume of a 20 nm nanopillar. In general, the mechanical property of the surface layer must be considered when explaining the deformation behavior of such a nano composite.

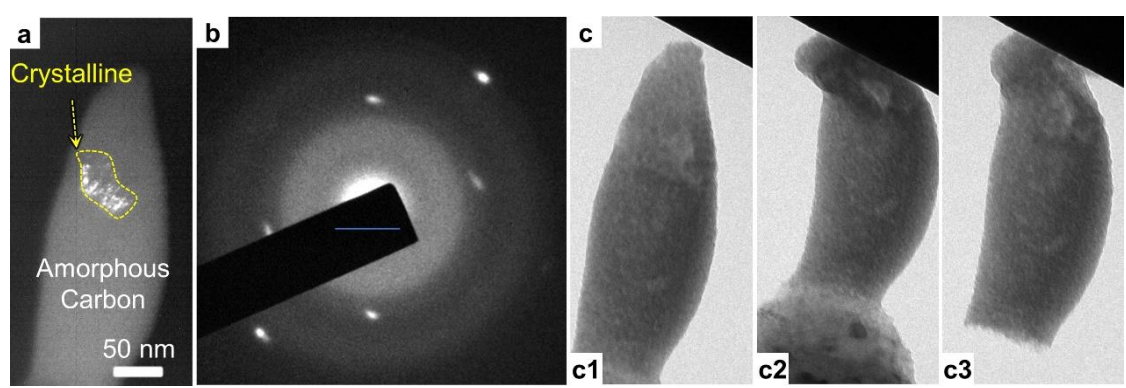

**Figure 3.** (a) Dark-field TEM image of an amorphous carbon nanopillar. Circled region with bright contrast contains a crystalline grain. The amorphous carbon displays uniform grey contrast. (b) The corresponding SAED pattern. The diffuse rings are typical for amorphous carbon. The sharp spots come from the crystalline grain. (c) Snapshots of the amorphous carbon

nanopillar (c1) before brought into contact with, (c2) when pushed against and (c3) after released from the indenter.

The *in-situ* compression in Fig. 3 shows that plastic deformation can be easily accommodated in nanoscale amorphous carbon. Therefore, we suggest that the permanent deflection in Fig. 2a3 was mainly induced by the plastic deformation of the amorphous layer. It is not clear whether this is the cause of the observation in Regan *et al.*'s work for lack of no atomic scale characterization of the nanopillar surface structure. [4] In a recent work by Banerjee *et al.*, [2] RIE-fabricated diamond nanopillars were observed to have amorphous surface layers with thickness ~ 2 – 3 nm by high-resolution TEM imaging. More information on the nanopillar characterization is necessary to exclude the possible surface effect on the deformation of diamond nanopillars with diameters less than 20 nm.

To elucidate the mechanism of plastic deformation observed in <001>-oriented nanopillars, MD simulations were performed on diamond undergoing stretching and bending deformation in Regan *et al.*'s work. [4] The Environment-Dependent Interaction Potential (EDIP) [6] was employed to predict the plastic deformation mechanisms. It is noted that the EDIP was developed to simulate the growth of amorphous carbon thin films based on the amorphous structures generated from melts at 5000 K [6]. Accordingly, the EDIP is most appropriate for simulating amorphous carbon instead of its crystalline form. [7] To the best of our knowledge, no dislocation structure has been predicted by this potential. Meanwhile, the diamond nanopillar in simulations has a diameter of 3 nm, which may not have large enough activation volume for dislocation nucleation. Consequently, dislocation slip may be completely excluded for the chosen interatomic potential and the MD model size. We also note that phase transformation highly depends on both thermodynamics and kinetics of the particular system under loading. A thermodynamically stable phase in MD simulations, even as validated by first-principle calculations, may not appear in experiments for the different kinetic paths. The presence of O8 carbon needs to be experimentally verified. The only experimental result provided by the authors is the electron diffraction pattern of an ensemble of diamond nanopillars (Fig. S5 in Supporting Information of [4]). It was claimed that the extra diffraction ring agreed well with the O8 phase (113) plane. However, it is hard to determine a new phase based on one diffraction ring because there are numerous possibilities for other phases containing the lattice plane with similar *d*-spacing. In addition, more experimental evidence, especially high-resolution TEM characterization, is necessary to exclude some byproduct phases or impurities potentially introduced during the sample fabrication, *e.g.* RIE-induced new phases or impurities in the CVD grown diamond.

In conclusion, our *in-situ* TEM study reveals the room-temperature plasticity can be carried out by dislocation slip in both <100> and <111>-oriented diamond nanopillars. The brittle-to-ductile transition is highly dependent on the stress state. For *in-situ* bending tests, cleavage fracture is dominant due to the presence of positive resolved normal stress on {111} planes. Under compression, crack nucleation and propagation are effectively suppressed due to the large hydrostatic stress component. Dislocations are activated by the resolved shear stress on the slip planes. However, our observations do not rule out other possibilities of room-temperature plasticity in nanoscale diamond. As suggested by Regan et al., [4] the plastic deformation may also be triggered by phase transformation in the nanopillars that are sufficiently small to induce ductile rather than a brittle deformation for the Gibbs Free energy to be lowered. In both studies, we note that the surface structure may play a significant role in the deformation mechanisms, as the incipient plasticity always occurs from the surface region in nanoscale diamonds.

**Acknowledgments:**

This work was supported by the National Natural Science Foundation of China (Grant No. 11725210, 11672355 and 11702165).

**Declaration of Interests:**

The authors declare no competing interests.

**References**

[1]  Q. Huang, D. Yu, B. Xu, W. Hu, Y. Ma, Y. Wang, Z. Zhao, B. Wen, J. He, Z. Liu, Nature 2014, 510, 250.

[2]  A. Banerjee, D. Bernoulli, H. Zhang, M.-F. Yuen, J. Liu, J. Dong, F. Ding, J. Lu, M. Dao, W. Zhang, Y. Lu, S. Suresh, Science 2018, 360, 300.

[3]  J. Field, C. Freeman, Philosophical Magazine A 1981, 43, 595.

[4]  B. Regan, A. Aghajamali, J. Froech, T. T. Tran, J. Scott, J. Bishop, I. Suarez-Martinez, Y. Liu, J. M. Cairney, N. A. Marks, M. Toth, I. Aharonovich, Advanced Materials 2020, n/a, 1906458.

[5]  A. Nie, Y. Bu, P. Li, Y. Zhang, T. Jin, J. Liu, Z. Su, Y. Wang, J. He, Z. Liu, H. Wang, Y. Tian, W. Yang, Nature Communications 2019, 10, 5533.

[6]  N. Marks, Phys. Rev. B 2000, 63.

[7]  C. de Tomas, I. Suarez-Martinez, N. A. Marks, Carbon 2016, 109, 681.